\documentclass[journal]{IEEEtran}

\usepackage{xcolor,soul,framed} 

\colorlet{shadecolor}{yellow}
\usepackage[pdftex]{graphicx}
\graphicspath{{../pdf/}{../jpeg/}}
\DeclareGraphicsExtensions{.pdf,.jpeg,.png}

\usepackage[cmex10]{amsmath}
\usepackage{array}
\usepackage{mdwmath}
\usepackage{mdwtab}
\usepackage{eqparbox}
\usepackage{url}

\begin{document}
\bstctlcite{IEEEexample:BSTcontrol}
    \title{Beam Steering by Two Vertically Faced Metasurfaces Using Polarization-Free Unit Cells with Three Operating Modes}
  \author{Mohammadreza Jahangiri,~\IEEEmembership{Student Member,~IEEE,}
      Hossein Soleimani,~\IEEEmembership{Member,~IEEE,}\\

       Mohammad Soleimani,~\IEEEmembership{Fellow,~IEEE}

  \thanks{}
  \thanks{}
  \thanks{}%
  \thanks{}
  \thanks{}}

\markboth{
}{Mohammadreza Jahangiri \MakeLowercase{\textit{et al.}}:Beam Steering by Two Vertically Faced Metasurfaces Using Polarization-Free Unit Cells with Three Operating Modes}

\maketitle

\begin{abstract}
Intelligent reflecting surfaces (IRS) are valuable tools for enhancing the intelligence of the propagation environment. They have the ability to direct EM Waves to a specific user through beamforming. 
A significant number of passive elements are integrated into metasurfaces, allowing for their incorporation onto various surfaces such as walls and buildings. In certain situations, metasurfaces may need to be installed on surfaces that are not flat, such as surfaces with curves or surfaces with two sides.  
In this work, a unit cell with three polarization modes, consisting of absorption, reflection, and a 1-bit phase shift, is designed and investigated. The unit cell in the nonplanar metasurface is composed of two vertical surfaces. A mathematical model is used to analyze two vertically-faced metasurfaces. The results include S parameter values and full structure. It is demonstrated that both absorption reflection mode and phase shift mode are achieved within a specific frequency band. The simulations demonstrate that an appropriate absorption level, reaching -10 dB, is achieved in absorption mode. In reflection mode, a 180-degree phase shift is achieved within the same frequency band.  Reports on the radiation pattern of two vertically faced structures indicate that there is no quantized beam when there is a one-bit phase shift.

\end{abstract}


%
\IEEEpeerreviewmaketitle


\section{Introduction}
\label{sec:introduction}
\PARstart{R}{econfigurable}  
Intelligent surfaces have gained significant interest due to their ability to control electromagnetic waves in wireless propagation environments. This is particularly relevant in the context of 5G and 6G wireless networks, where the small wavelength of millimeter waves makes them susceptible to attenuation and scattering by small objects. As a result, intelligent surfaces play a crucial role in allowing the propagated waves to bypass obstacles and efficiently focus on the desired user. With RIS, the propagation environment can be controlled alongside the transmitter and receiver. Beamforming using RIS enables precise beam direction and coverage control for users. The reconfigurable feature of the metasurface allows for dynamic beam steering. 

The Multiple Input Multiple Output (MIMO) technique improves network capacity and tackles challenging propagation environments. The beamforming technique in massive MIMO was initially introduced in the 4G Network  \cite{b1}. Numerous studies have explored the use of hybrid beamforming in the context of massive MIMO and have been classified accordingly \cite{b2}. The number of antenna elements in the Massive MIMO method can increase implementation costs and complexity. This is due to the presence of an RF chain block for each antenna element. The RIS Concept involves the use of passive elements instead of active antenna elements for the implementation of massive MIMO.\\

Phase shift is a crucial component of beamforming. Using a phase gradient on a metasurface, it becomes possible to break conventional Snell's law and precisely steer the reflected ray to the desired angle. In conventional Snell's law, the absence of a phase gradient results in the reflected wave angle being the same as the incident wave angle. Most metasurfaces utilize a discrete phase shift. For example, pin diodes \cite{b3}, MEMS switches \cite{b4}, mechanical adjustments \cite{b1}, and changes in chemical potential \cite{b5} can be used to tune metasurface and reflect array unit cells to make phase shifts of 1 bit or 2 bits. This allows for dynamic changes in the phase state of each unit cell. The general concepts of IRS were introduced in \cite{b6}. Metasurfaces are crucial in improving the signal-to-noise ratio in user positions within 5G cellular networks. This is achieved by carefully adjusting the state of each cell, leading to optimal performance \cite{b7}. Each cell state expresses the cell phase shift and the cell attenuation derived by a designed unit cell. For example, previous studies \cite{b8,b9} have utilized different techniques, such as using pin diodes connected to the ground plane in a shunt connection. \cite{b10} employed current reversal techniques, and \cite{b11} focused on developing a dual-polarized, two-bit reflect array unit cell that utilizes eight-pin diodes. The interfaces and design of the unit cell are crucial factors, surpassing the significance of lumped elements. A resonator is utilized at millimeter-wave frequencies to minimize the impact of pin diode fluctuations on phase shift, as demonstrated in \cite{b12}. One disadvantage of the 1-bit metasurface is the generation of a quantized beam, which is absent in the 2-bit metasurface. One possible solution to this issue is incorporating an offset phase into the metasurface to suppress the unwanted beam \cite{b13}. \\
In most applications, additional functions are often required alongside phase shifting. These functions may include a transmission-to-reflection switch using a pin diode \cite{b14}, integrated circuit \cite{b15}, moving dielectric slab \cite{b1}, and changing the chemical potential in a graphene-based metasurface \cite{b16}. In addition to transmission and reflection, absorption function is added to the metasurface in reference \cite{b17}. Single-polarized works typically use one or two diodes \cite{b3,b18}, while dual-polarized works require multiple pin diodes in each metasurface cell \cite{b10} for symmetrical structure. Most symmetrical structures exhibit four-angle shapes, such as cross \cite{b10} or ring \cite{b19}. The symmetrical structure can also be achieved by using an equilateral triangle structure, which is not affected by polarization \cite{b20}. The mentioned metasurface designs were confined to planar metasurfaces. There will be a need to study metasurfaces in two vertically-faced, cylindrical, or spherical shapes. The study of wave behavior over curvilinear metasurface is demonstrated in \cite{b21} by analytical approach.

This paper introduces a metasurface with a 3-mode function that includes absorption, reflection, and phase shift. The metasurface operates in dual polarization by utilizing a symmetrical structure with three-pin diodes positioned symmetrically for each layer. The bottom layer is responsible for controlling absorption and reflection, while the top layer is responsible for controlling phase shift. The simulation results for the unit cell and full structure design at 26.8 GHz are provided in the next part. The simulation results demonstrate two metasurface sheets that are vertically faced and applicable in wireless networks. These designs utilize the designed unit cell to demonstrate the superior performance of the quantized beam in a 1-bit structure positioned between two vertical planes.

\section{Design and principle}
\label{sec:Design}
In this section, we will discuss the design and theoretical issues of the unit cell, as well as the theoretical analysis of the two-faced metasurface. Two steps are involved in estimating the MIMO channel in a wireless network \cite{b22}. During the initial stage of channel estimation, it is necessary to switch off the RIS to estimate the direct channel between the base station and the user. In the second stage, each cell is switched on to estimate the channel for each cell. RIS-Assisted Communication considers two propagation paths: the RIS path and the direct path. The RIS unit cells must operate in both absorption and phase shift modes. This paper presents the design of a RIS operating in three different modes. The primary principle of line-of-sight beamforming metasurface is to change the reflecting angle through phase change. The generalized form of Snell's law, as shown in Equation \eqref{eq1}, demonstrates how the direction of the reflected wave is changed by the phase gradient of the metasurface structure.

\begin{figure*}
    \centering
    \includegraphics[width=4.5in]{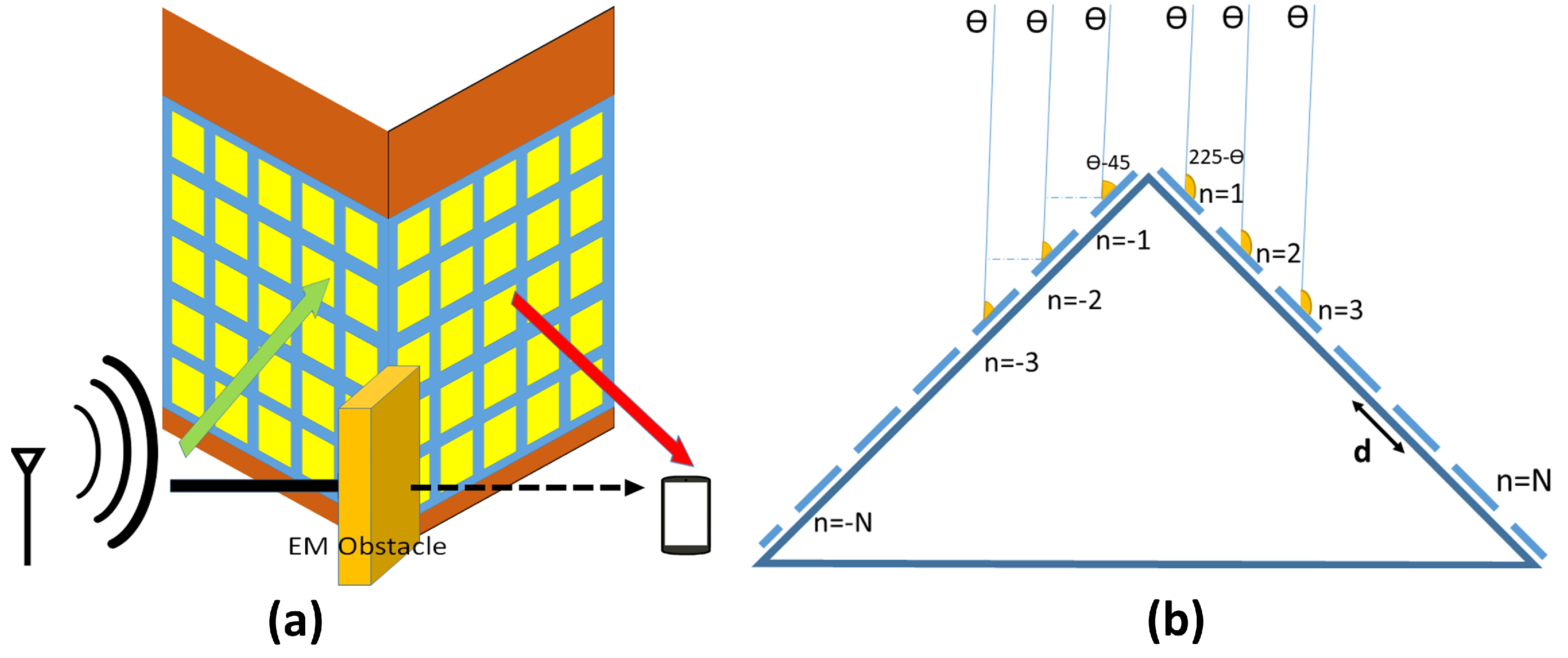}
    \caption{Illustration of two metasurfaces faced vertically. (a) How it navigates around obstacles (b) Geometry of two linear arrays with equal spacing of $d$ between array elements and an incident wave angle in the $\theta$ direction.}
    \label{fig1}
\end{figure*}

\begin{equation}
Sin({\theta _r}) - Sin({\theta _i}) = \frac{{{\lambda _0}}}{{2\pi {n_i}}}\frac{{d\phi }}{{dy}}
\label{eq1}
\end{equation}

The phase condition of each cell is determined by the difference between the induced phase by the incident wave and the required phase to steer the beam to the desired angle.The continuous phase of each unit cell is calculated using Equation \eqref{eq2}. The calculated phase is quantized to its closest value of 0 or 180 degrees according to Equation \eqref{eq3} in the 1-bit meta surface. The calculated phase in a 2-bit metasurface is quantized to the nearest value of 0, 90, 180, or 270 degrees.

\begin{equation}
{\varphi _{nm}} = {\varphi _r}\left( {{x_n},{y_m}} \right) - {\varphi _i}\left( {{x_n},{y_m}} \right)
\label{eq2}
\end{equation}
where:

${\varphi _r}\left( {{x_n},{y_m}} \right) =  - jksin\left( {{\theta _r}} \right)[{x_n}\cos \left( {{\varphi _r}} \right) + {y_n}\sin \left( {{\varphi _r}} \right)]$

and 

${\varphi _i}\left( {{x_n},{y_m}} \right) =  - jksin\left( {{\theta _i}} \right)[{x_n}\cos \left( {{\varphi _i}} \right) + {y_n}\sin \left( {{\varphi _i}} \right)]$

\begin{equation}
{\alpha _{1bit,nm}} = \left\{ {\begin{array}{*{20}{c}}
0&{ - \pi /2 \le {\varphi _{nm}} \le \pi /2}\\
\pi &{Otherwise}
\end{array}} \right\}
\label{eq3}
\end{equation}
Where $\theta$ is  wave elevation angle and $\phi$ is  wave azimuth angle radiation pattern synthesize by metasurface can be approximated by using reflect array patch antenna equation shown in Equations \eqref{eq4} and \eqref{eq5}: 
\begin{equation}
\scalebox{1}{$\begin{array}{*{20}{c}}
{{E_r}\left( {\theta ,\varphi } \right) = {\rm{cos}}\left( \theta  \right)\mathop \sum \limits_{n,m = 1}^N {\Gamma_{mn}}{E_i}\left( {{x_n},{y_n}} \right){\rm{cos}}\left( {{\theta _{nm}}} \right)}\\
{ \times {\rm{exp}}( - jksin\left( \theta  \right)\left[ {{x_n}\cos \left( \varphi  \right) + {y_n}\sin \left( \varphi  \right)} \right]}
\end{array}$}
\label{eq4}
\end{equation}
Where
\begin{equation}
{\Gamma_{mn}} = {\alpha _{mn}}\;{\rm{exp}}\left( { - j{\alpha _{1bit,nm}}} \right)
\label{eq5}
\end{equation}
\subsection{Two vertically faced array}
In certain wireless communication scenarios, it is necessary to utilize two vertical faces of a metasurface for beamforming, as using just one face is not enough to cover the desired area. The concept of a two-faced metasurface is illustrated in Figure 1. Two metasurfaces are embedded on two vertical faces. The faces can potentially serve as walls for a building.

 Equations \eqref{eq6} and \eqref{eq7} show the array factor of two linear arrays positioned vertically, as shown in Figure 1(b).
\begin{equation}
\begin{array}{*{20}{c}}
{{\rm{F}}\left( {\rm{\theta }} \right) = \mathop \sum \limits_{n =  - N}^{n = 0} {e^{jnkdcos\left( {45 - \theta } \right) - {\varphi _n}}} + }\\
{\mathop \sum \limits_{n = 1}^{n = N} {e^{ - jkndCos\left( {45 - \theta } \right) + jknd\sqrt 2 Cos\left( \theta  \right) - {\varphi _n}}}}
\end{array}
\label{eq6}
\end{equation}
\begin{equation}
\scalebox{0.85}{${\alpha _n} = \left\{ {\begin{array}{*{20}{c}}
{jnkdcos\left( {45 - {\theta _0}} \right)}&{n \le 0}\\
{ - jkndCos\left( {45 - {\theta _0}} \right) + jknd\sqrt 2 Cos\left( {{\theta _0}} \right)}&{n > 0}
\end{array}} \right\}$}
\label{eq7}
\end{equation}
The extension of Equations \eqref{eq6} and \eqref{eq7} for 2D planar Array is given as \eqref{eq8} \eqref{eq9}.
\begin{equation}
\scalebox{1}{$\begin{array}{*{20}{c}}
{{\rm{F}}\left( {{\rm{\theta }},\phi } \right) = \mathop \sum \limits_{m = 0}^{m = M} \mathop \sum \limits_{n =  - N}^{n = 0} {e^{jnkxcos\left( {45 - \varphi } \right) + mkysin\left( \theta  \right) - {\varphi _{mn}}}} + }\\
{\mathop \sum \limits_{m = 0}^{m = M} \mathop \sum \limits_{n = 1}^{n = N} {e^{ - jknxCos\left( {45 - \varphi } \right) + jknx\sqrt 2 Cos\left( \varphi  \right) + mkysin\left( \theta  \right) - {\varphi _{mn}}}}}
\end{array}$}
\label{eq8}
\end{equation}

\begin{equation}
\scalebox{0.8}{${\varphi _{mn}} = \left\{ {\begin{array}{*{20}{c}}
{\begin{array}{*{20}{c}}
{jnkdcos\left( {45 - {\varphi _0}} \right)\; + mkysin\left( {{\theta _0}} \right)}\\
{}
\end{array}}&{n \le 0}\\
{\begin{array}{*{20}{c}}
{ - jkndCos\left( {45 - {\varphi _0}} \right) + jknd\sqrt 2 Cos\left( {{\varphi _0}} \right)}\\
{ + mkysin\left( {{\theta _0}} \right)}
\end{array}}&{n > 0}
\end{array}} \right\}$}
\label{eq9}
\end{equation}

Where $\theta$ represents the  beam elevation angle, $\phi$ is the beam azimuth angle, and $\phi_{mn}$ is the phase shift of elements $m$ and $n$. $d$ represents the spacing of two adjacent elements, as shown in Figure 1(b). Equations \eqref{eq7} and \eqref{eq9} show the phase shift required for the 1D and 2D linear array. Equation \eqref{eq8} is used to calculate the induced phase caused by the incident wave that arrives at angle $\theta_i$ based on Equation \eqref{eq10}.

\begin{equation}
\scalebox{0.85}{${\varphi _{inducted,m}} = \left\{ {\begin{array}{*{20}{c}}
{jmkdcos\left( {45 - {\theta _i}} \right)}&{m \le 0}\\
{ - jkmdCos\left( {45 - {\theta _i}} \right) + jkmd\sqrt 2 Cos\left( {{\theta _i}} \right)}&{m > 0}
\end{array}} \right\}$}
\label{eq10}
\end{equation}
The phase required for each meta surface element to steer the beam towards the desired direction $\theta_r$ is provided in Equation \eqref{eq11}. 

\begin{equation}
\scalebox{0.85}{${\varphi _{Reflected,m}} = \left\{ {\begin{array}{*{20}{c}}
{jmkdcos\left( {45 - {\theta _r}} \right)}&{m \le 0}\\
{ - jkmdCos\left( {45 - {\theta _r}} \right) + jkmd\sqrt 2 Cos\left( {{\theta _r}} \right)}&{m > 0}
\end{array}} \right\}$}
\label{eq11}
\end{equation}

Where $m$ is the index of meta surface element. The phase shift required to steer the incident angle $\theta_i$ to reflection angle $\theta_r$ is given in Equation \eqref{eq12}.

\begin{equation}
{\varphi _{PhaseShift,m}} = {\varphi _{reflected,m}} - {\varphi _{inducted,m}}
\label{eq12}
\end{equation}

 For a 1-bit metasurface, the phase shift calculated in Equation \eqref{eq12} must be quantized according to Equation \eqref{eq13}.
\begin{equation}
{\varphi _{1bit,m}} = \left\{ {\begin{array}{*{20}{c}}
0&{0 \le {\alpha _{PhaseShift,m}} \le 180}\\
{180}&{Otherwise}
\end{array}} \right\}
\label{eq13}
\end{equation}

\subsection {3 mode Unit cell design} 
Figure 2. shows the main geometry and functions of the reconfigurable unit cell operating in three modes. The lower layer controls absorption and reflection, while the top layer controls the phase shift. The layer beneath the ground plane is a DC control layer that does not have any impact on the other layer functions. Table 1 displays the upper and lower pin diode states during the operation of these three modes. Due to their rapid switching capability, millimeter wave pin diodes (MACOM-000907) are utilized for switching absorption mode to reflection mode in the lower layer and achieving a 180-degree phase difference in the top layer when the lower layer is in reflection mode. The MACOM-000907 pin diode has an equivalent circuit of $L=0.05nH$ and $R=4.2\Omega$ in the on state, and $L=0.05nH$ and $C=42fF$ in the off state \cite{b23}.  \\
\begin{figure*}
    \centering
    \includegraphics[scale=.12]{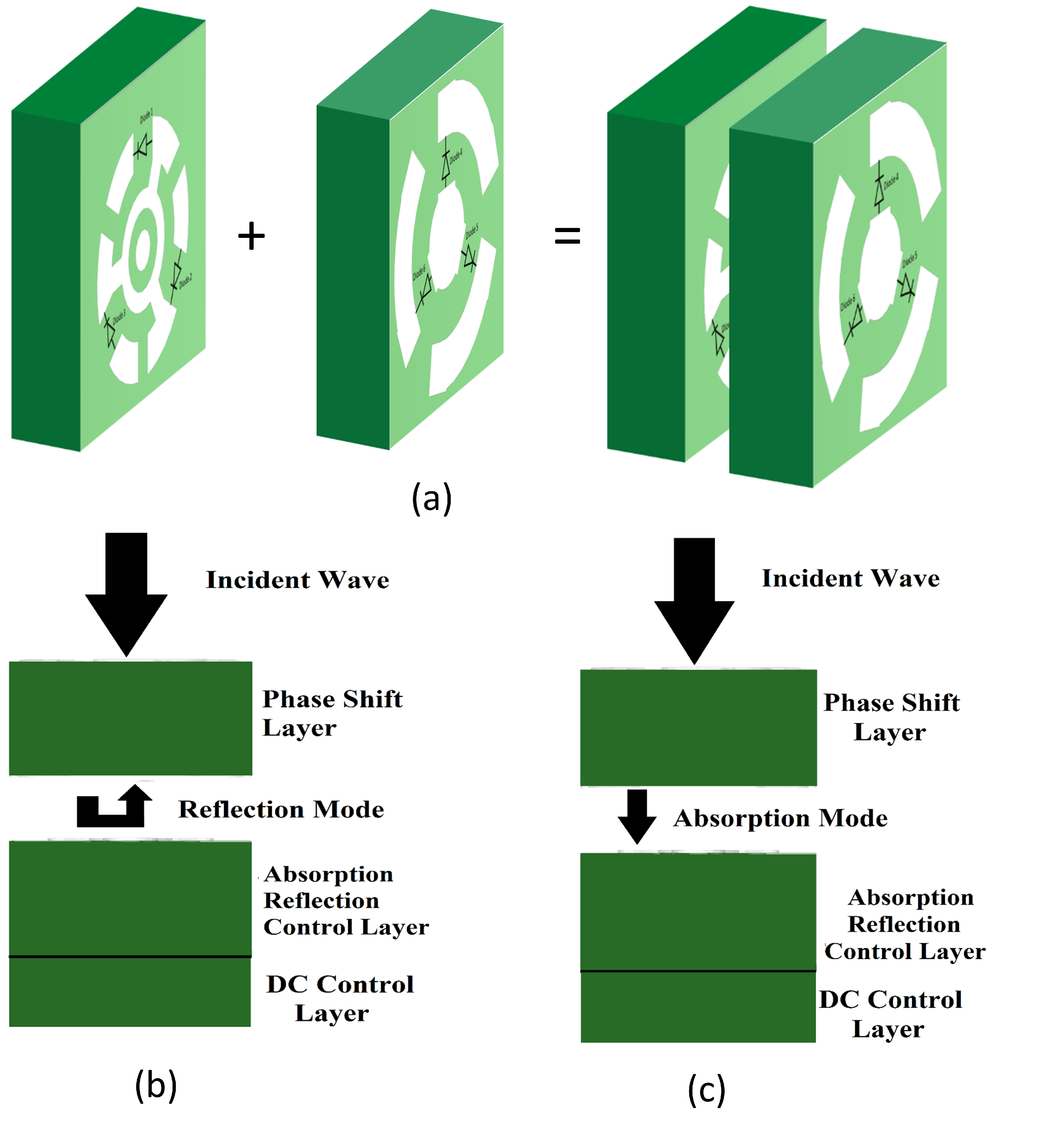}
    \caption{ The geometry of the 3-mode metasurface and layer functions. (a) General concept of two layer positions and functions; (b) Reflection mode; (c) Absorption mode.}
    \label{fig2}
\end{figure*}
Since the unit cell is polarization-independent, it is necessary for the unit cell structure to be symmetric. In most studies, a quad-angle structure is utilized to achieve polarization independence. However, in this study, a three-angle structure is introduced in order to minimize the number of pin diodes and reduce costs. Three diodes are positioned symmetrically, as shown in Figure 2(a). The detailed geometry is shown in Figure 3, alongside the parameter values provided in Table 2. 
\begin{table}[]
\centering
\caption{states of pin diodes in 3 mode operating work}
\begin{tabular}{llll}
  & Mode                       & Upper PIN diodes   & Lower PIN diodes   \\ \hline
1 & Absorption                 & 15 mA Forward bias & 15 mA Forward bias \\ \hline
2 & Reflection 0 phase shift   & 15 mA Forward bias & -5 V Reverse bias  \\ \hline
3 & Reflection 180 phase shift & -5 V Reverse bias  & -5 V Reverse bias  \\  \hline
\end{tabular}
\end{table}
\subsection{Lower layer design }
The suggested design configuration in \cite{b19} includes a split ring absorber.  The design of a reconfigurable absorber split ring is shown in Figure 3(c), while the corresponding equivalent circuit is depicted in Figure 4. When the upper layer is in forward bias, it becomes transparent to the incident wave. By placing pin diodes in the gaps, switching from absorption mode to reflection mode is achieved by electronically changing the resonance frequency of the split ring. The resonance frequency is obtained from Equations \eqref{eq14} and \eqref{eq15}. 
\begin{equation}
{F_{res}} = \frac{1}{{\sqrt {L{C_{total}}} }}
\label{eq14}
\end{equation}

\begin{equation}
{C_{total}} = C + {C_{var}}
\label{eq15}
\end{equation}

\begin{figure*}
    \centering
    \includegraphics[scale=0.18]{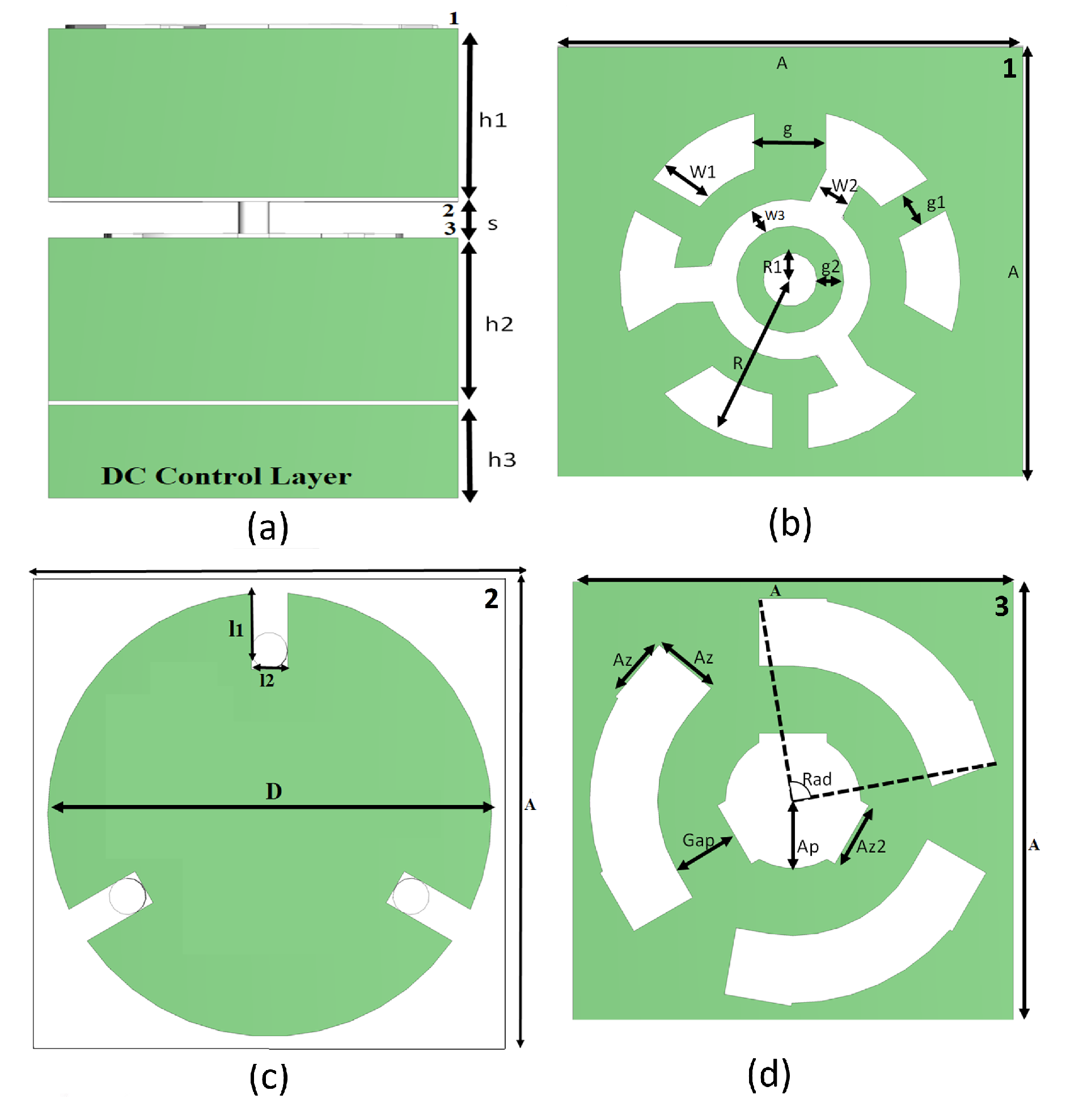}
    \caption{Schematic of the unit cell. (a) Side view. (b) Bottom part of the upper layer. (c) Bottom layer. (d) Top part of the upper layer. The positions of (b), (c), and (d) are illustrated in (a).}
    \label{fig3}
\end{figure*}
\begin{table*}[]
\centering
\caption{Design parameters given in Figure 3.}
\begin{tabular}{llllll}
A   & 2.6mm   & W2 & 0.2 mm  & h1 & 1.4     \\ \hline
Az  & 0.4mm   & W3 & 0.15mm  & h2 & 1.35    \\  \hline
Ap  & 0.45 mm & g  & 0.4 mm  & h3 & 0.77 mm \\    \hline
Gap & 0.4 mm  & R  & 0.95 mm & S  & 0.3 mm  \\    \hline
Az2 & 0.4 mm  & R1 & 0.15 mm & D  & 2.45mm  \\     \hline
Rad & 65 deg  & g1 & 0.2 mm  & l1 & 0.42mm  \\     \hline
W1  & 0.3 mm  & g2 & 0.15mm  & l2 & 0.2mm   \\     \hline
\end{tabular}
\end{table*}

Where L and C represent the equivalent capacitance and inductance of the split ring. By changing ${C_{total}}$, the resonance frequency is changed, and the absorption mode is transferred to another frequency. The split ring gap is modeled as a capacitor, while the ring and other parts are modeled as inductors. Applying DC voltage to the pin diode parallel circuit changes the gap capacitor and has an impact on the resonant frequency. Figure 3(c). shows the design geometry of the lower layer. Another gap is utilized for the purpose of isolating DC bias between the anode and cathode sides of pin diodes. The fr4 substrate is commonly used for lower layer design in high-loss scenarios, particularly at millimeter wave frequencies. It has a loss tangent of $\sigma$=0.032 and $\epsilon$ =4.15 at 22-28 GHz, and the copper thickness is $35\mu m$. Figure 3. illustrates the impact of substrate thickness on resonance frequency in absorption mode. As the thickness increases, the absorption frequency decreases. A thickness of 1.35 mm for fr4 is selected for the h2 parameter because it resonates at 26.8 GHz. The smaller values of h2 result in a reduced reflection loss at the absorption frequency. The anode side of the pin diode is connected to the control layer of the metasurface. The cathode side of the pin diode is connected to the ground using vias with a diameter of 0.2mm. The central part of the design connects the DC control layer to the phase shift layer.

\begin{figure}

    \centering
    \includegraphics[scale=0.09]{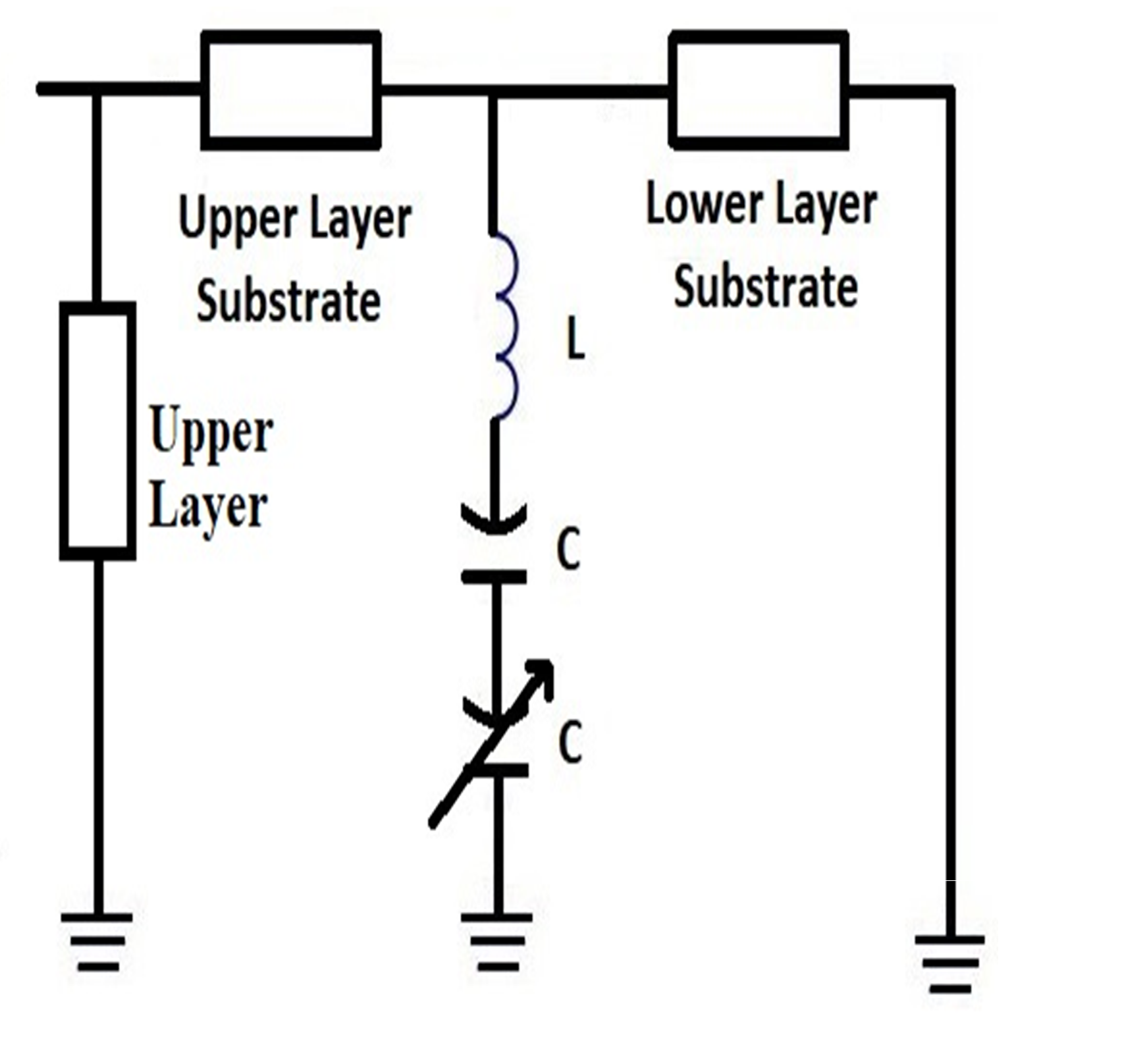}
    \caption{ Equivalent circuit of the designed unit cell. The bottom layer is split ring with coil and capacitor equvalent circuit, while the top layer is patch and three delay lines act transparently when the top layer pin diode are in forward bias mode.}
    \label{fig4}
\end{figure}

\begin{figure}

    \centering
    \includegraphics[scale=0.05]{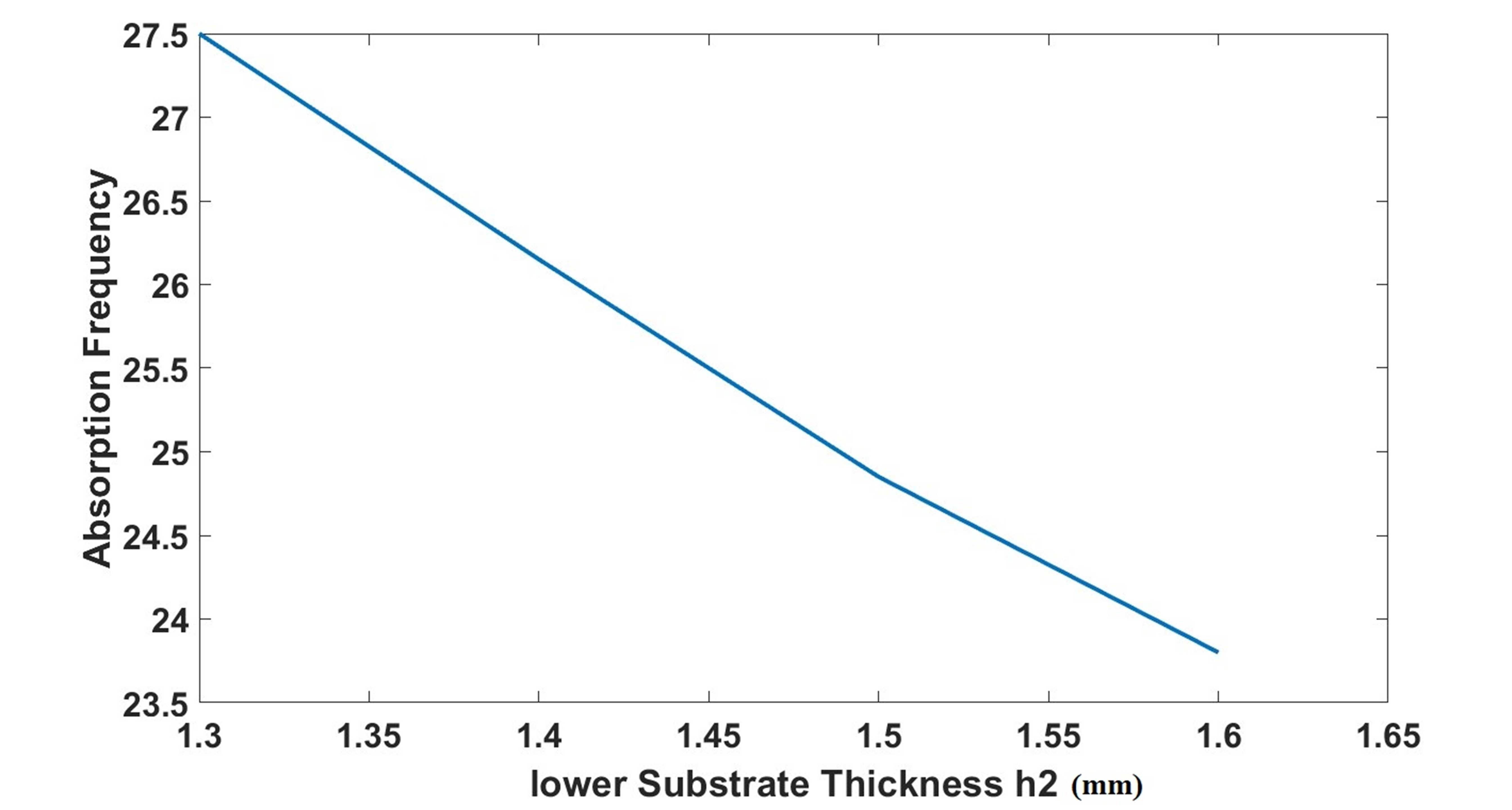}
    \caption{ Effects of the fr4 substrate thickness (h2) on the absorption resonance frequency. When pin diodes are in state 1, as shown in Table 1.}
    \label{fig5}
\end{figure}

\subsection{Upper layer design} 	
  The upper layer is responsible for controlling phase shift. The main patch connection to three external parts is controlled through the use of biasing pin diodes. The additional component includes a delay line that is connected to the ground through vias. The substrate used in the upper layer is RT duroid 5880, with a $\epsilon_r=2.2$ and a dielectric loss tangent of $\sigma $=0.009. Vias behave similarly to short circuits at low frequencies, while having a coil-like function at high frequencies. Phase shifting is implemented by the delay line, and via connects it to ground. The schematics of the upper layer are depicted in Figure 3(d), while the corresponding parameter values can be found in Table 2. Figure 6 illustrates the impact of variations in the Rad variable from Table 2 on phase shifting. As shown in Figure 2, an air gap is present between the phase shift control layer and the reflection absorption control layer. This gap allows for the placement of lower-layer pin diodes. 
\begin{figure}
    \centering
    \includegraphics[scale=.05]{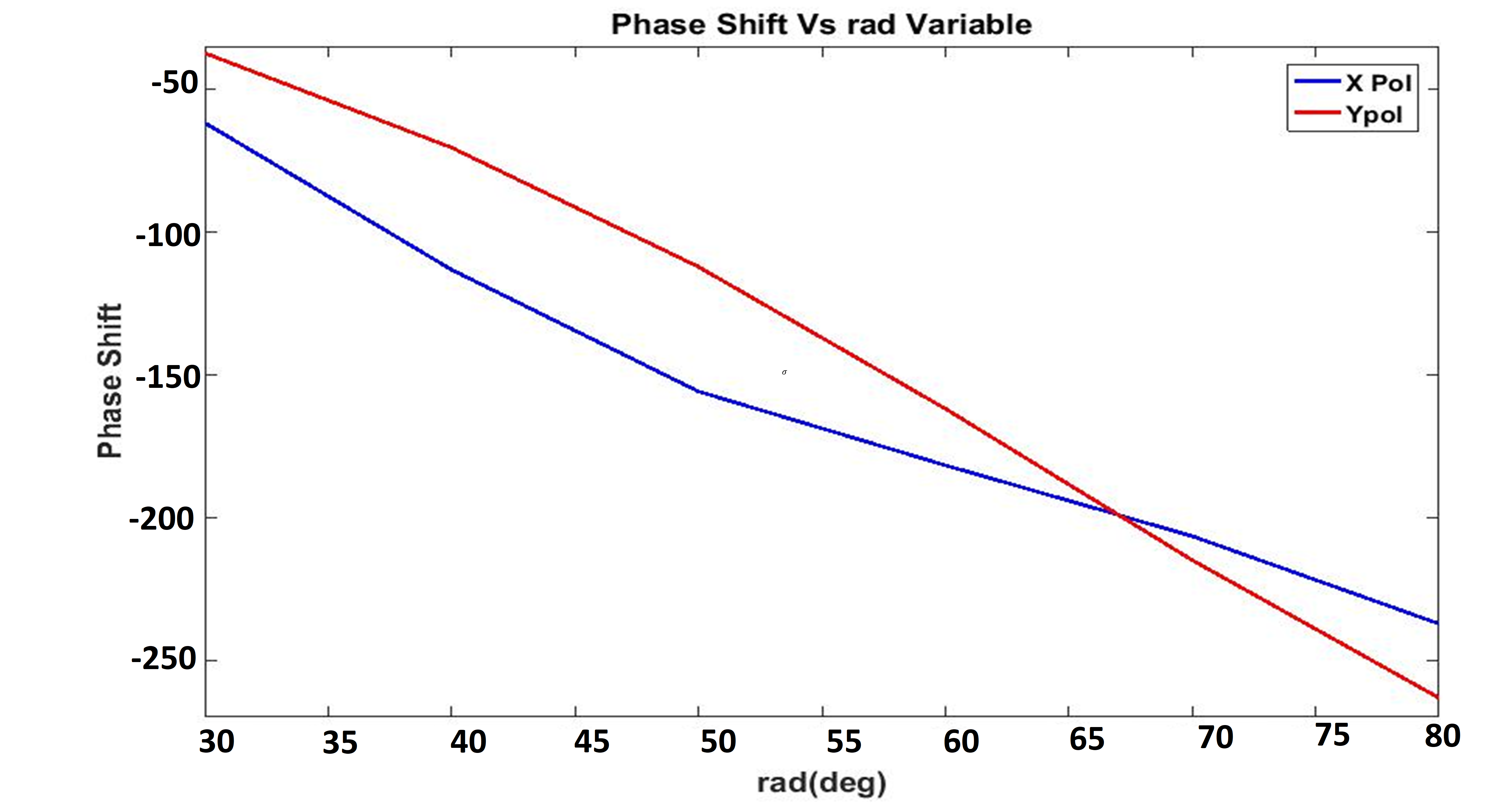}
    \caption{ Illustration of the changes in phase shift in relation to the variations in the radial size of the delay line (represented by the ‘rad’ variable in Figure 3(d)). It is observed that a radial size of 65 degrees facilitates a 180-degree phase shift, in addition to a minimal phase shift difference in both x and y polarizations.}
    \label{fig6}
\end{figure}

\section{Simulation Results}
Simulation and results
In some cases, IRS can be embedded on non-planar surfaces to bypass obstacles in wireless propagation environments, as discussed in the previous section. This section presents the simulation results of the unit cell and the full structure design of a planar metasurface and two vertically facing metasurfaces. The full structure results are shown for a planar metasurface with a size of 16*16 and two vertically faced metasurfaces with a size of 14*14. The absorption frequency for a normal split ring is 32 GHz, while adding an inner ring and vias decreases the absorption frequency to 26.8 GHz. The absorption and phase shift results of the designed unit cell are shown in Figure 7 for both x and y polarizations. It is important to note that the polarization conversion of the unit cell remains consistently below 19.7 dB across all frequencies, as shown in Figure 7(a). 
\begin{figure}
    \centering
    \includegraphics[scale=.12]{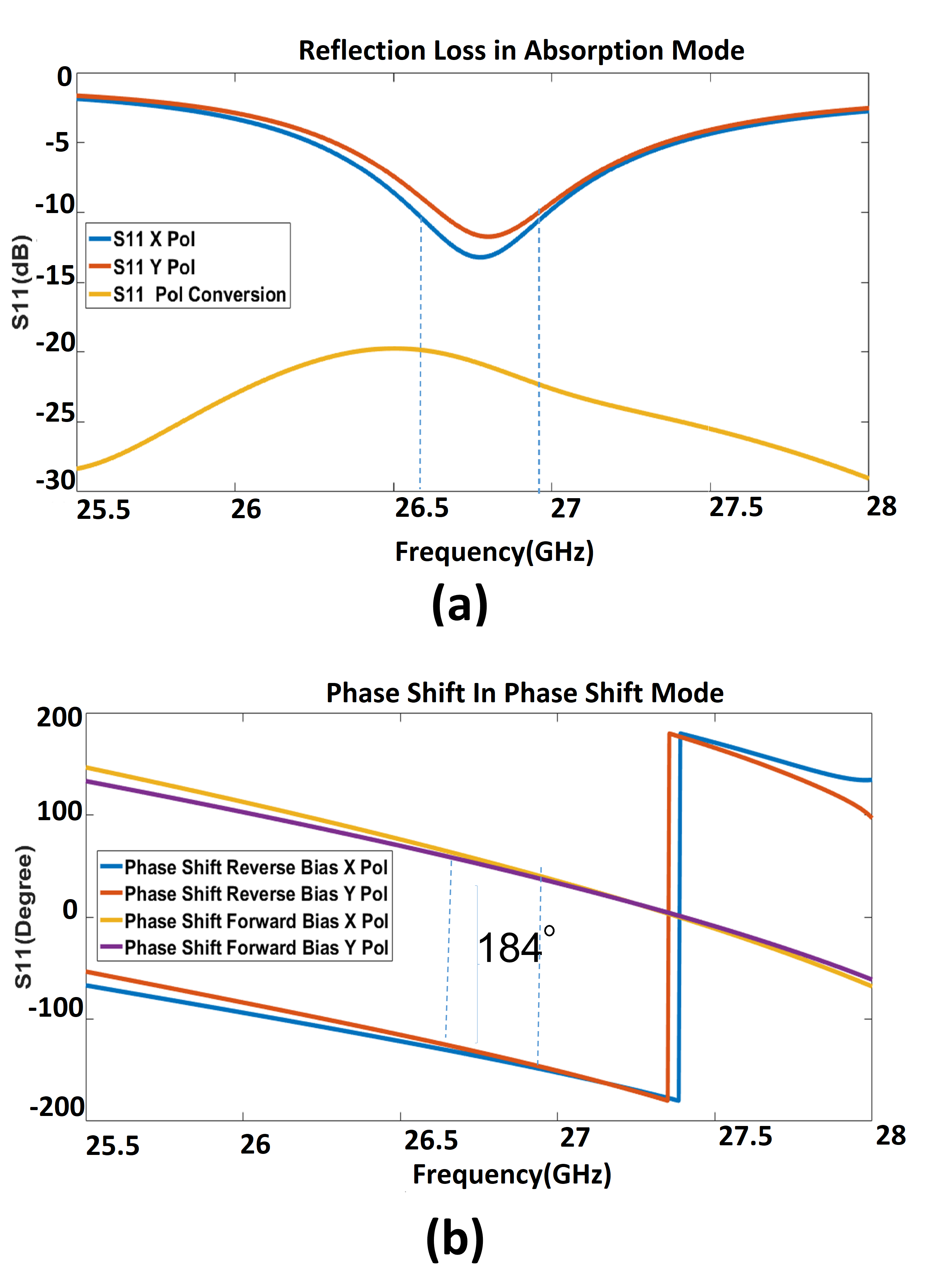}
    \caption{ Unit cell simulation results: (a) S-parameter in absorption mode, and (b) Phase shift in reflection mode for x and y polarizations.}
    \label{fig7}
\end{figure}

Figure 7(a) shows that a S parameter below -10 dB is achieved in the frequency band of 26.5–27 GHz. Figure 7b demonstrates a significant phase shift of 184 degrees in reflection mode across a wide range of frequencies, specifically 26.5–27 GHz.As discussed in the previous section, utilizing two vertical faces can be beneficial when the coverage area is located near a corner of a wall or building. This allows for beamforming to involve both faces. The designed unit cell is utilized in the full structure design. The geometry of the full structure design is shown in Figure 8(a) and Fig8(b). It is important to mention that the beamforming target should be located within -45 to +45 of the array broadside in order to involve both faces of the metasurface in two-faced metasurface beamforming. If the beamforming target is outside the -45 to +45 range, only one of the two vertically facing metasurfaces is used for beamforming, while the target is in the dead zone of the other face.  
\begin{figure}
    \centering
    \includegraphics[scale=.16]{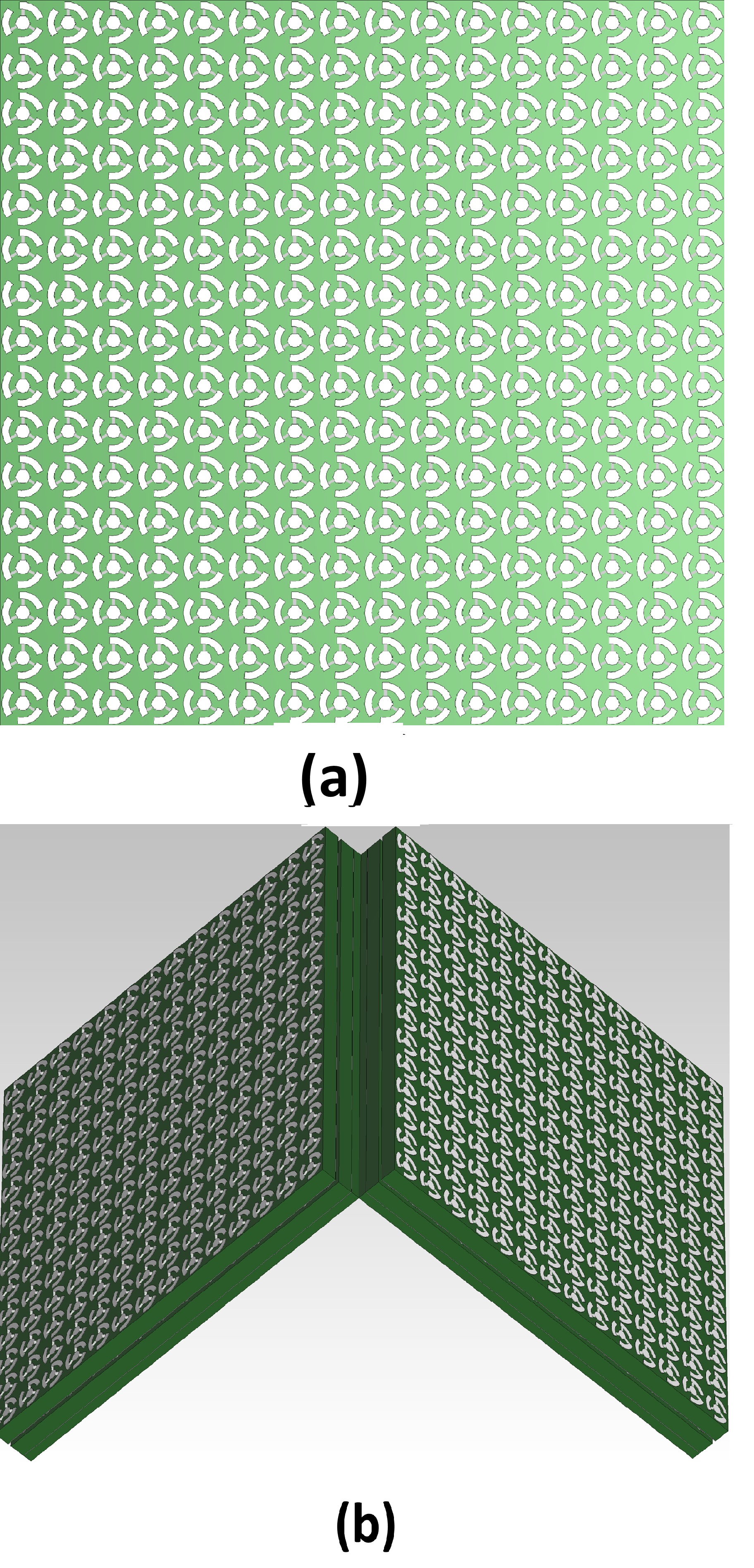}
    \caption{ Schematics of the full-structure metasurface: (a) Planar, and (b) Two vertically-faced metasurfaces.}
    \label{fig8}
\end{figure}

The radiation pattern of the planar metasurface in reflection mode is shown in Figure 9 at a frequency of 26.8 GHz. It demonstrates three different reflection angles for the three metasurface supercell states when the incident wave comes from the array broadside. The unit cells' phase shift configuration for each radiation pattern is shown below the radiation pattern. The smaller size of the supercell results in a greater steering angle.

\begin{figure*}
    \centering
    \includegraphics[scale=.12]{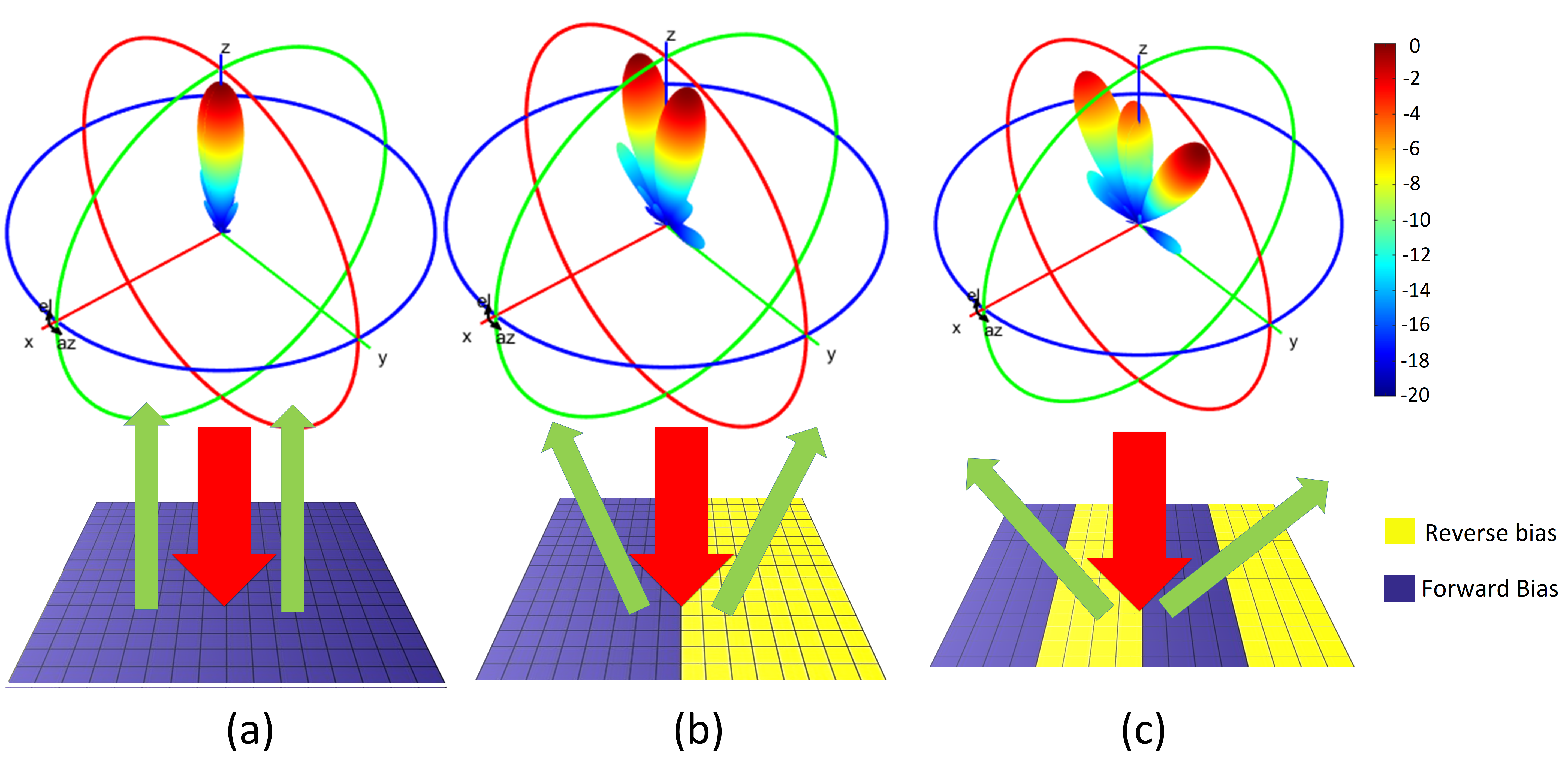}
    \caption{ Radiation pattern of the planar metasurface at 26.8 GHz with different phase configurations for incident angles from the broadside.(a) no super cell leads to no steering direction ;(b)8 size supper cell leads to 12 degrees steering angle ;(c)4 size supper cell leads to 32 degrees steering angle ; }
    \label{fig9}
\end{figure*}

The phase shift configuration of two vertically faced metasurfaces is shown in Figure 10 for three reflection angles when the incident wave is coming from the broadside. It is observed that the quantization beam does not exist in the two-faced metasurface, even with a one-bit phase shift. A quantized beam is generated in a planar metasurface when there is a similarity between the phase shifts of 0 and 180 degrees, as well as -180 and 0 degrees, when the inducted phase is equal in all cells. In the case of a two-faced metasurface, the equality of these two phase states is disrupted because of the varying phase shift induced on the unit cells. This can be seen in Equation \eqref{eq16}, which is another representation of Equation \eqref{eq12}.
\begin{equation}
{\varphi _{Reflected,m}} = {\varphi _{Quantized,m}} + {\varphi _{inducted,m}}
\label{eq16}
\end{equation}
Figure 10 shows the radiation pattern of both the full structure simulation and the analytical model for a one-bit metasurface in a vertically faced geometry at 26 GHz. This frequency band effectively achieves a 180-degree phase difference, resulting in a desirable radiation pattern. the radiation pattern results are illustrated for both full wave simulation and MATLAB simulation.

\begin{figure*}[hbt!]
    \centering
    \includegraphics[scale=0.12]{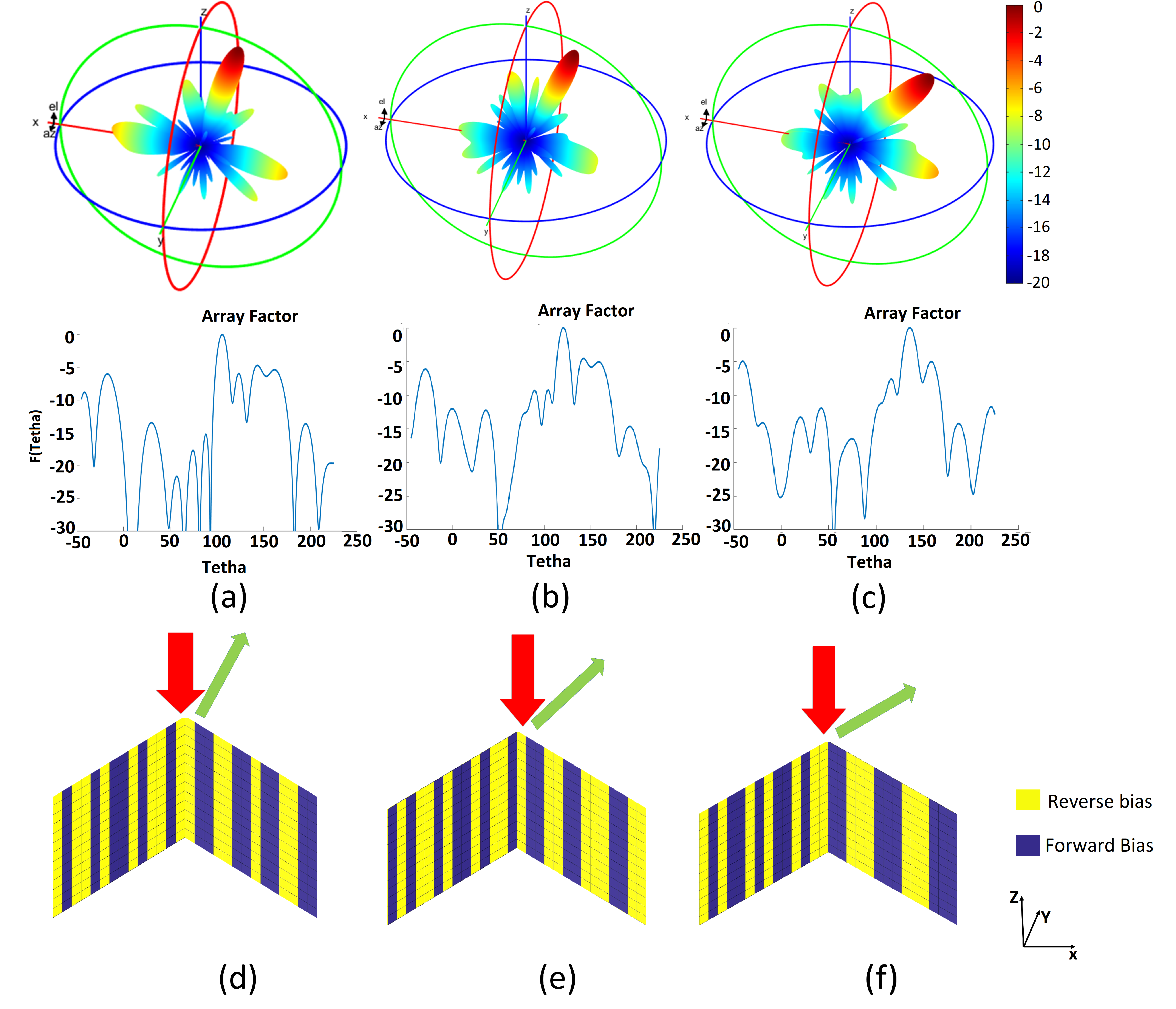}
\caption{ Radiation pattern of two vertically-faced metasurfaces at 26 GHz obtained in full wave simulation(3D radiation pattern), alongside the analytical model of the radiation pattern obtained in MATLAB simulation for ideal 180 degrees phase shift(2D radiation pattern). The incident wave originates from the broadside and is demonstrated in red arrows. with desired reflection direction of (a) 15 degrees; (b) 30 degrees;(c) 45 degrees. The unit cell phase states are demonstrated for 2 vertical faces for (d) 15 degrees reflection direction; (e) 30 degrees reflection direction; (f) 45 degrees reflection direction.}
    \label{fig10}
\end{figure*}

An analytical model is shown in Figures 10(g), 10(h), and 10(i), showcasing two vertically faced metasurfaces consisting of 2*14*14 passive elements. It explores reflection angles of 15, 30, and 45 degrees, as described by Equations \eqref{eq6} and \eqref{eq7} from the previous section. The incident wave is assumed to approach from the broadside. The results are provided for azimuth angles ranging from -45 to 225 degrees. These angles correspond to the area in front of the metasurface, similar to the range of 0 to 180 degrees for a planar metasurface. In practical massive MIMO systems, the number of passive elements embedded on a giant surface is significantly large, resulting in a narrow beam.

\section{Conclusion}
In wireless communication, metasurfaces need to be embedded on various surfaces and should not be affected by polarization.  This paper investigates the two vertically faced metasurfaces using three-mode polarization-free unit cells. The symmetrical geometry of the unit cell is achieved through a combination of a 3-angle structure and pin diodes. The simulation results demonstrate that there is absorption occurring in the frequency range of 26.5–27 GHz in absorption mode, while in reflection mode, there is a phase shift of 180 degrees within the same frequency band. The full structure simulation on a planar surface and two vertically faced surfaces is conducted. The results demonstrate that effective beam steering is achieved through the appropriate distribution of phase shifts in metasurface unit cells.

\ifCLASSOPTIONcaptionsoff
  \newpage
\fi

\

\end{document}